\documentclass[reprint,amsmath,twocolumn,amssymb, nobibnotes, aps, pre, superscriptaddress]{revtex4-1}

\usepackage{graphicx}
\usepackage{dcolumn}
\usepackage{bm}
\usepackage{color}
\usepackage{tikz}

\begin{document}
\def \beq{\begin{equation}}
\def \eeq{\end{equation}}
\def \bea{\begin{eqnarray}}
\def \eea{\end{eqnarray}}
\def \bem{\begin{displaymath}}
\def \eem{\end{displaymath}}
\def \P{\Psi}
\def \Pd{|\Psi(\boldsymbol{r})|}
\def \Pds{|\Psi^{\ast}(\boldsymbol{r})|}
\def \Po{\overline{\Psi}}
\def \bs{\boldsymbol}
\def \bl{\bar{\boldsymbol{l}}}
\newcommand{\ihat}{\hat{\textbf{\i}}}
\newcommand{\jhat}{\hat{\textbf{\j}}}

\newcommand{\eps}{\varepsilon}

\title{{Transverse Instability of Rogue Waves}} 
\author{Mark J. Ablowitz}
\affiliation{Department of Applied Mathematics, University of Colorado, Boulder, Colorado {80309,} USA}
\author{Justin T. Cole}
\affiliation{Department of Mathematics, University of Colorado, Colorado Springs, Colorado {80918,} USA}
\date{\today}     

\begin{abstract}
Rogue waves are abnormally large waves which appear unexpectedly and have attracted considerable attention, {particularly in recent} years. The  one space{,} one time (1+1) 
nonlinear Schr\"odinger 
equation is 
often used to model rogue waves;
it is an envelope description of 
plane wave{s} and admits 
the so-called Pergerine and Kuznetov-Ma  
soliton solutions.
{However,} {in deep water waves and certain electromagnetic systems where there are two significant transverse dimensions,} the 2+1 
hyperbolic 
nonlinear Schr\"odinger equation 
is the appropriate wave envelope description. 
Here we show that {these} 
rogue wave solutions suffer from 
strong transverse instability {at long {\it and} short frequencies. Moreover, the stability of the Peregrine soliton is found to coincide with that of the background plane wave. These results indicate that, {when} applicable, transverse dimensions {must} be taken into account when {investigating} rogue wave pheneomena.} 

\end{abstract}
\maketitle 


%

In recent years researchers have studied a class of large amplitude waves that  {were} 
previously 
relegated to folklore{:} 
giant {water} waves appearing out of nowhere causing extreme damage to, and even loss of, {maritime vessels.} 
The first {verified} measurement of an extreme water wave was Jan 1, 1995 on  the Draupner platform in the North Sea  
 where a 25.6m (84ft) wave was observed \cite{Draup2004}{; 
 {much larger than} the 
 {background wave field}. Subsequently, rogue waves events have been observed in several laboratory settings such as wave tanks \cite{Chaub2011,Chaub2013}, nonlinear optics \cite{Solli2007,Kibler2010,Lecaplain2012}, superfluid helium \cite{Ganshin2008}, and plasmas \cite{Bailung2011}.} 
 
 The 1+1 dimensional nonlinear Schr\"odinger (NLS) 
 equation is a {well-known} model {used to describe} 
 the 
 {envelopes} 
 of 
 {generic} 
 nonlinear plane wave solutions {in one spatial dimension (corresponding to the direction of propagation) and one temporal dimension {\cite{YangBook}}. In} water waves{, the 
  nonlinear plane wave} 
 {was} found by G. Stokes in the mid-1800's {\cite{Stokes1847}. {Importantly,} the} 
  {underlying} 
 NLS equation admits special large amplitude {``rogue wave"} solutions {including} the Peregrine \cite{Per1983} and  Kuznetov-Ma (KM) solitons \cite{Kuz1977,Ma1979}. 
 However, a more accurate wave envelope approximation of deep water waves takes into account the transverse variation; this is the 2+1 dimensional NLS equation --see Eq.~(\ref{sNLS}) below. 
 
{A} {physically significant} regime in deep water waves corresponds to small surface tension, which is described by the {2+1 `hyperbolic' NLS  (HNLS) equation \cite{MJAHS1981}. Another important application of the HNLS equation is 
spatiotemporal electromagnetc 
wave propagation in 
media with normal dispersion {cf.} \cite{MJA2011,Trapani2003}}.  {While other more sophisticated models do exist,
NLS models can suggest possible mechanisms that will be relevant in applications even where NLS  might not be the {optimal} model.} {We note that other 2+1 water wave models} 
are frequently used in the study of rogue waves{,} such as crossing states 
\cite{Oranato2006,MJA2015,Bremer2019}{, but} we do not discuss {those} 
phenomena here. 
 {The KM and Peregrine 
  solitons are known to be unstable in 1+1 dimensions {for scalar} \cite{Cuevas2017,Calini2019} {and vector systems \cite{Baronio2014}},
but  {{corresponding} stability analysis in 2+1 dimensions  has not been carried out.}}
In this case we show that both {of these solitons}  
suffer from transverse instability.

{Transverse 
instability of nonlinear solutions is an important and well-known effect, e.g., the transverse instability of localized solitons {by} 
long wavelengths 
was first found 
in the mid 1970s \cite{Zak1974}.} 
{By} employing Floquet theory we show  that there are  instabilities at arbitrarily small transverse wavelength scales with finite growth rates.  Indeed{,} it has been long known that
the Stokes plane wave solution {of} 
the HNLS equation {is} 
unstable to transverse variations \cite{Zak1968,Benney1969,DS1974}.
With this observation it is perhaps to be expected that the Peregrine and  Kuznetov-Ma solitons{,} which {at large distances} limit to the plane wave{,}  will also suffer from this serious instability. But what is remarkable is the similarity {of the instability profiles} between the {plane waves and  
KM/Peregrine solitons, particularly at high transverse frequencies{.} {{Furthermore,} in the hyperbolic case these rogue waves are found to have an instability region for {\it all} transverse wave numbers, as opposed to a finite region instability 
{like} 
the elliptic version.}} 

We 
note 
a major difference between the {instabilities of the} 
elliptic 
{and} 
hyperbolic NLS {equations.} 
The elliptic version does 
not have growth rates at arbitrarily small transverse wavelengths; this instability, first found in one dimensional water waves \cite{BF1967}, is often termed modulational instability \cite{Zakharov2009}. {Moreover, the elliptic-focusing NLS equation with cubic nonlinearity can suffer from collapse in finite time \cite{Vlasov1970} whereas{, to our knowledge, it is not known if the hyperbolic version 
exhibits finite time wave collapse}.}  {Numerical evidence that Peregrine solitons can collapse in the elliptic NLS equation was found in \cite{Klein20}.}


{{The} governing NLS equation is introduced and the relevant soliton solutions are presented {below}. 
The stability of plane wave solutions and KM solitons {are} 
calculated: the former by 
analytic methods and the latter 
numerically through Floquet theory. 
Direct numerical simulations confirm 
our stability findings. We conclude that a full 2+1 study is {important} for rogue wave systems {when there are two significant} transverse dimensions.}

 {Consider a plane wave envelope $u(x,y,t)$  propagating through a nonlinear dispersive media  in a preferred $x$-direction. 
 Going to a translating coordinate frame moving at the 
 group velocity and then 
 changing variables 
leads to the dimensionless 
 NLS equation 
{\begin{equation}
iu_t+u_{xx}+s_1u_{yy} + 2s_2 \left( |u|^2  - u_0^2 \right) u=0 ,
\label{sNLS}
\end{equation}
} where $s_1, s_2 =\pm 1$. As $|x| \rightarrow \infty$, $u$ approaches {a constant, $u_0$;}
~without loss of generality, we set $u_0 =  1$. In water waves, the sign of the coefficients} 
depends on the surface tension
--cf. Fig.~4.15 in \cite{MJAHS1981}. 
For small surface tension (ST) we have: $s_1=-1, s_2=1$, termed {hyperbolic;} 
for moderate ST: $s_1=1, s_2=-1$, {called} elliptic-defocusing; {and at} sufficiently large ST: $s_1=s_2 = 1$, {termed} elliptic-focusing. {The hyperbolic-focusing ($s_1=-1,s_2=1$) and hyperbolic-defocusing ($s_1=-1,s_2=-1$) equations are equivalent up to conjugation and exchange of $x$ and $y$ in (\ref{sNLS}). As a result, they will yield similar stability results below and we shall simply refer to the ``hyperbolic-focusing'' signs as the ``hyperbolic'' NLS equation from here on.}

\begin{figure}
\centering
\includegraphics[scale=.32]{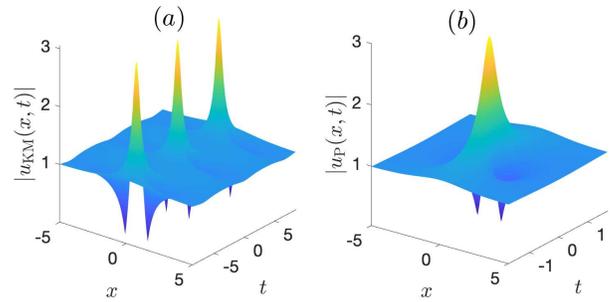}
\caption{Magnitude evolutions of the (a) Kuznetzov-Ma soliton in (\ref{KM}) with $Z = 1.25$ and (b) Peregrine soliton in (\ref{P}). \label{Soliton_plots}}
\end{figure}

{There are two well-known {1+1} soliton solutions of the 
focusing NLS equation ($s_2 = 1$)  we shall focus on.
The first is the 
KM soliton, 
which is periodic in $t$ and localized in $x$,  given by 
\begin{align}
\label{KM}
u_{\rm KM}&(x,t) = \\ \nonumber
& \frac{\cosh c_{-}x+\frac{1}{2c_{+}}(c_{+}^2+c^2_{-})\sin s-ic_{-}\cos s}{\cosh c_{-}x +\frac{2}{c_{+}}\sin s} ,
\end{align}
where  $s=c_{+}c_{-} t - \pi/2 $, $ c_{\pm}=Z\pm1/Z$, $Z > 1$ [see Fig.~\ref{Soliton_plots}(a)]. The parameters are chosen so that the soliton peak  occurs at $x = 0$ and the largest (smallest) magnitude $ |u_{\rm KM}(0,t)|$  at times $t = n T$ {$\left(\frac{(2n+1)T}{2} \right)$} {for} $n \in \mathbb{Z}${,} with period $T = 2 \pi / (Z^2 - 1/Z^2)$.}
As {$Z \to 1$}, this solution approaches the Peregrine soliton 
\begin{equation}
u_{\rm P}(x,t)=  \frac{4x^2-16it+16t^2-3}{4x^2+16t^2+1},
\label{P}
\end{equation}
which is localized in both $x$ and $t$ since $T \rightarrow \infty$ [see Fig.~\ref{Soliton_plots}(b)]. At large distances {both} 
solutions approach {a} 
plane wave, i.e., {as} $|x| \to \infty$, {$u_{\rm KM},u_{\rm P} \to \exp(2 i t)$.} 
In terms of the inverse scattering transform, we assume that as $|x| \to \pm \infty, |u|$ tends  to unity sufficiently {fast}{. This} 
class includes the KM soliton (with exponentially fast decay), but 
not 
the Peregrine soliton {which decays algebraically fast} cf. \cite{Bion2014}.

{We now study the stability of these solitons to transverse perturbations. Consider the two-dimensional perturbation {
$$u(x,y,t) =   \widetilde{u}(x,t) + w(x,y,t) $$ 
} where {$ \widetilde{u}(x,t)   $} is a solution of (\ref{sNLS}) and the function $w $ has small magnitude. Linearizing (\ref{sNLS}) about the solution gives 
\begin{equation}
iw_t+w_{xx}+s_1w_{yy} + 2s_2\left[ \left( 2 |\widetilde{u}|^2 - 1 \right)w+ \widetilde{u}^2 w^* \right]=0 ,
\end{equation}
where $w^*$ is the complex conjugate of $w$. We look for Fourier solutions of the form 
\[ w(x,y,t) = w_{+}(x,t)e^{i \ell y}+w_{-}(x,t)e^{-i \ell y} . \] 
Setting the coefficients of $e^{\pm i \ell y}$ to zero yields
\begin{align}
\nonumber
&i\partial_t w_{+} +[ \partial_x^2 -s_1\ell^2 + 2s_2 (2 |\widetilde{u}|^2 - 1) ] w_{+}+2s_2 \widetilde{u}^2 w_{-}^*=0 \\
\label{stab_system}
&i\partial_t w_{-}^* -[ \partial_x^2 -s_1\ell^2+2s_2 (2 |\widetilde{u}|^2 - 1)] w_{-}^*-2s_2 (\widetilde{u}^*)^2w_{+}=0 .
\end{align}
{We} consider two cases: ({\it a}) modulational instability (MI) of a plane wave where {{$\widetilde{u}(x, t) =1 $}} and $w_\pm(x,t) = \alpha_{\pm } \exp[\pm i ( k x - \lambda t) ]$; and ({\it b}) transverse instability of the KM soliton where {$\widetilde{u}(x,t)   =  u_{\rm KM}(x,t)$} and $w_{\pm}$ is a Floquet mode which satisfies 
\begin{equation}
\begin{pmatrix}
w_+ \\ w_-^*
\end{pmatrix}
(x,t + T ) = e^{- i \lambda (\ell) T} \begin{pmatrix}
w_+ \\ w_-^*
\end{pmatrix}(x,t)
\end{equation}
{for} 
Floquet exponent $\lambda(\ell)$ {\cite{Eastham73}}. 

For the Peregrine soliton i.e. {$\widetilde{u}= u_{\rm P}$}, the form of the eigenfunction $w$ is unclear since the coefficients in (\ref{stab_system}) are localized in both $x$ {\it and} $t$. Instead we take an indirect approach: since the KM soliton approaches the Peregrine soliton as $Z \rightarrow 1$, we  {expect} 
that the stability of KM will also approach that of {Peregrine.}}

\begin{figure}
\centering
\includegraphics[scale=.29]{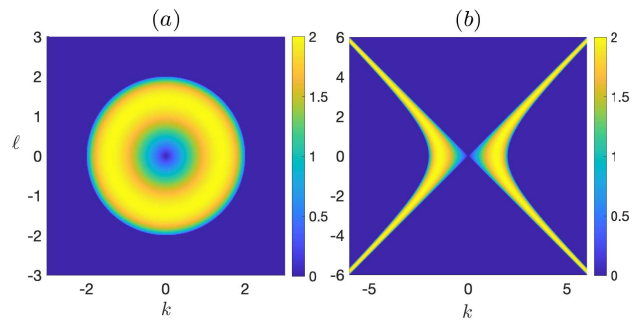}
\caption{{Plane wave {instability (MI)}:} (a) Elliptic-focusing ($s_1 = 1 = s_2$) and (b) hyperbolic 
($s_1=-1$ {, $s_2=1$}) instability regions and growth rates. Plotted is $|{\rm Im}\{ \lambda \}|$ for $\lambda(k,\ell)$  in (\ref{grwth}). \label{MI_stab_regions}}
\end{figure}

{For plane wave solutions, 
stability system (\ref{stab_system}) has constant coefficients. 
 Looking for plane wave eigenmodes  
 yields the eigenvalues}
 \begin{equation}
\lambda^2=(k^2+s_1\ell^2)(k^2 +s_1\ell^2 -4s_2)
\label{grwth}
\end{equation}
{which correspond to {linear} instability 
when ${\rm Im}\{\lambda\} \not= 0$.} 
When $\ell=0$ we recover the {classic one-dimensional MI result: unstable for $s_2=1$ (focusing) and stable when $s_2=-1$ (defocusing) {cf.} \cite{Zakharov2009}.} 

{The two-dimensional problem corresponds to 
$\ell \not= 0$ where} the plane wave is stable  in the elliptic-defocusing case: $s_1=1, s_2=-1$; {and}  in the elliptic-focusing case {
\[ s_1=s_2=1 \text{~~instability {occurs} when:~}  0 < k^2+\ell^2 <4 . \]
} {The {punctured} {disk} 
region of instability is shown in Fig.~\ref{MI_stab_regions}(a) with maximal instability of $\lambda_{\max} = \pm 2i $  along the circle $k^2 + \ell^2 = 2$.
This is reminiscent of the classic 1D MI result in that there is a band-limited region in the Fourier plane corresponding to unstable wavenumbers. Outside this long wavelength region{,} the linear stability analysis above does not predict {any} exponential growth.} 

{In} the {hyperbolic} 
{equation 
 (which corresponds to deep water waves with small ST) 
\[ s_1=-1, s_2=1 \text{~~instability {occurs} when~} \ell^2<k^2<\ell^2+4 . \]
The hyperbolic region of instability is shown in Fig.~\ref{MI_stab_regions}(b)  with largest instability of $\lambda_{\max} = \pm 2i $  along the hyperbola $k^2 - \ell^2 = 2$.}
We see that there is instability for {\it arbitrarily small wavelengths} {i.e. large $|k|,|\ell|$,}  with {a} finite growth rate when  $k^2=\ell^2+\gamma_1^2$:
\[ \lambda^2=\gamma_1^2(\gamma_1^2-4), ~~~~ 0<\gamma_1^2<4 . \]
Since the KM and Peregrine soliton limit {as $x \rightarrow \pm \infty$ is this} 
unstable plane wave{,} we expect serious instability will {also} ensue {for these solitons}. This is confirmed by {numerical calculations below. 
\begin{figure}
\centering
\includegraphics[scale=.4]{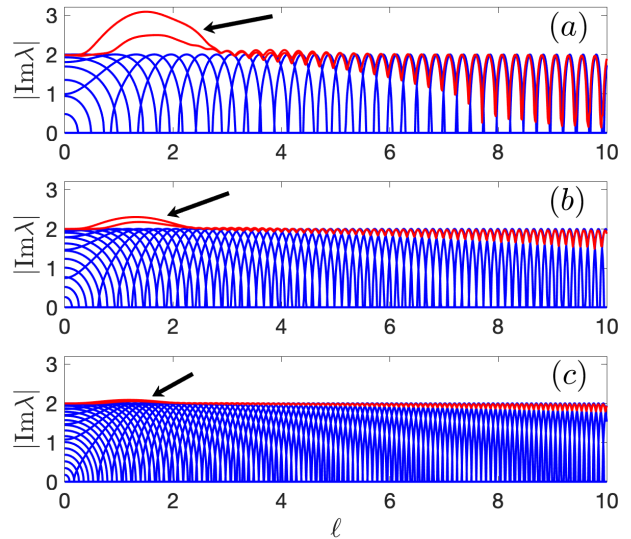}
\caption{{Hyperbolic 
($s_1=-1, s_2=1)$ {transversely unstable modes}; {blue curves: unstable plane wave} modes as function of transverse wavenumber $\ell$.  
 {red curves: two largest  unstable KM modes} {for}}  (a) $Z = 2$, (b) Z = 1.5,  (c) Z  = 1.25. Arrows point at the KM soliton instability curves. {At large $\ell$ values, the plane wave and KM soliton instability curves {nearly} overlap.}  
  \label{MI_KM_trans_stab}}
\end{figure}

The hyperbolic 
stability problem (\ref{stab_system}) is next solved with KM coefficients (\ref{KM}) at different transverse wavenumbers {using an exponential time-differencing integrator \cite{Kassam05}.} {We point out that the 1+1 results in \cite{Cuevas2017} correspond to the $\ell = 0$ case.} The imaginary (unstable) part of the Floquet exponents is shown in Fig.~\ref{MI_KM_trans_stab} {(red curves)}. Also shown are the unstable eigenvalues for plane wave solutions {(blue curves).}
{As}  $Z \rightarrow 1$, the decay rate approaches {that of} {the Peregrine soliton;} 
as a result, the computational window used to solve the problem is widened to ensure effectively {constant} 
boundary conditions for large $|x|$. One consequence of this is {additional} unstable modes and {faster} 
apparent rapidity of the humps in Fig.~\ref{MI_KM_trans_stab} {\cite{SuppNum}}. 

Overall, there is remarkable overlap between the soliton and plane wave instabilities, especially at large $\ell$. 
{An} important observation is that {\it KM solitons, like the background plane waves, are transversely unstable at high frequencies}. 
{As $Z \rightarrow 1$, the  spectrum of the KM soliton is approaching that of the plane wave; consisting of {rapidly varying} periodic hump-like structures bounded by 2.}
{It is remarkable that}  {\it the stability spectrum of a Peregrine soliton is the same as that of a plane wave.} {Our results suggest that the instability of rogue waves {for large transverse wave numbers can be conjectured based on the properties of the plane wave background.} }



\begin{figure}
\centering
\includegraphics[scale=.4]{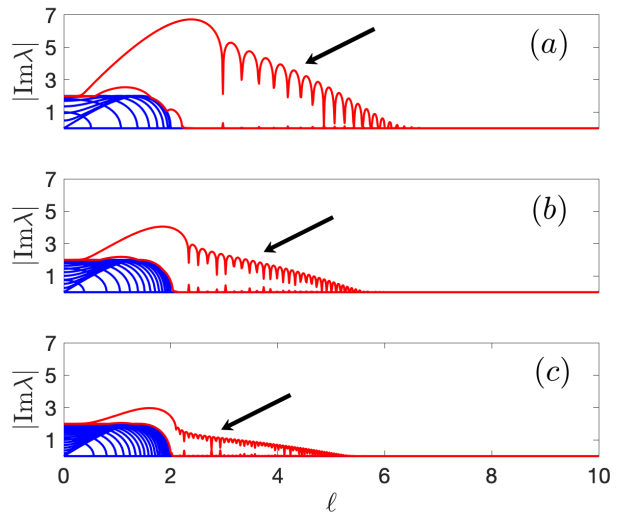}
\caption{{Elliptic-focusing ($s_1=1, s_2=1)$ {transversely unstable modes}; {blue curves: unstable plane wave} modes as function of transverse wavenumber $\ell$.  
 {red curves: two largest  unstable KM modes} {for}}  (a) $Z = 2$, (b) Z = 1.5,  (c) Z  = 1.25. Arrows point at the KM soliton instability curves. \label{Rogue_Wave_Instab_Elliptic}}
\end{figure}

The stability of the KM solitons in the elliptic-focusing NLS equation were also computed (strong ST regime in {deep} water waves). {The results, shown in Fig.~\ref{Rogue_Wave_Instab_Elliptic}, exhibit typical instability at long wavelengths. As $Z \rightarrow 1$, the interval of transversely unstable KM modes shrinks and the maximum magnitude {over} 
all $\ell$ 
tends to decrease and approach 2. Similar to the hyperbolic case, the soliton instability appears to be approaching {that of} the plane wave{,} 
though not as dramatically. Unlike the hyperbolic case{,} there is no  instability {indicated} for large $\ell${.}
{As indicated in Fig.~\ref{MI_KM_trans_stab} and Fig.~\ref{Rogue_Wave_Instab_Elliptic}{,} 
we see that in the hyperbolic case there is an instability regime for all transverse wave numbers whereas the instability zone for the elliptic case has finite size.}  



\begin{figure}
\centering
\includegraphics[scale=.36]{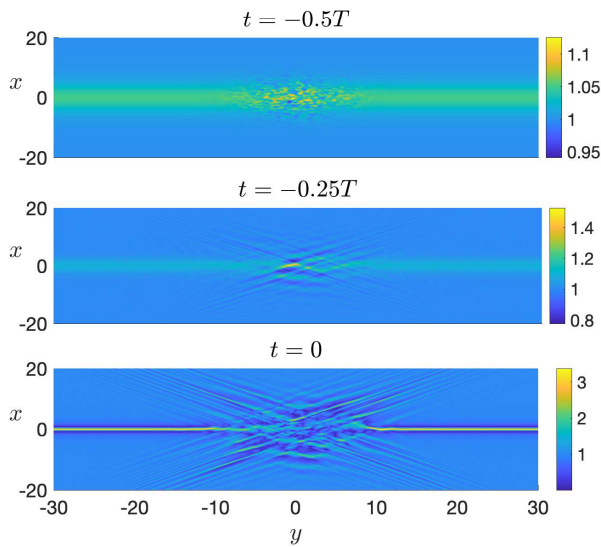}
\caption{Snapshots of {the} perturbed KM soliton {($\widetilde{u} = u_{\rm KM}$)} evolution ($Z = 1.25, T \approx 6.81$) {for} 
hyperbolic 
NLS ($s_1 = -1 = -s_2 $). Shown is $|u(x,y,t)|$ {seeded} with a $10\%$ localized perturbation at {$\overline{t} = - T/2$}.  \label{KM_DNS_Instab}}
\end{figure}

Finally, we examine the evolution of transversely perturbed rogue waves. Consider a perturbed solution  at time $t = \overline{t}$ of the form {
\begin{equation}
\label{IC_perturb}
u(x,y,\overline{t}) = \widetilde{u}(x,\overline{t}) +  w(x,y) ,
\end{equation}
} where $\widetilde{u}(x,\overline{t})$ is a line soliton solution and $w(x,y)$ is a normally distributed random function modulated by  a slowly decaying 
Gaussian function{. The peak magnitude of $w$ is taken to be 10\% that of {$\widetilde{u}(x,\overline{t})$} and typically we take $\overline{t} = -T/2$, where the KM soliton has minimal magnitude {\cite{SuppNum}}.}


First, consider the KM soliton at $Z = 1.25${; we recall it} has the instability spectrum shown in Fig.~\ref{MI_KM_trans_stab}(c). 
By the time the soliton reaches its peak magnitude at $t = 0$, the soliton has { disappeared} 
near the perturbation {and {an} `x-wave' has started to develop}{ -- see Fig. \ref{KM_DNS_Instab}}. 
We point out that this 
{is} the result of a perturbation whose magnitude is roughly 3\% the maximum {soliton} peak. 

\begin{figure}
\centering
\includegraphics[scale=.36]{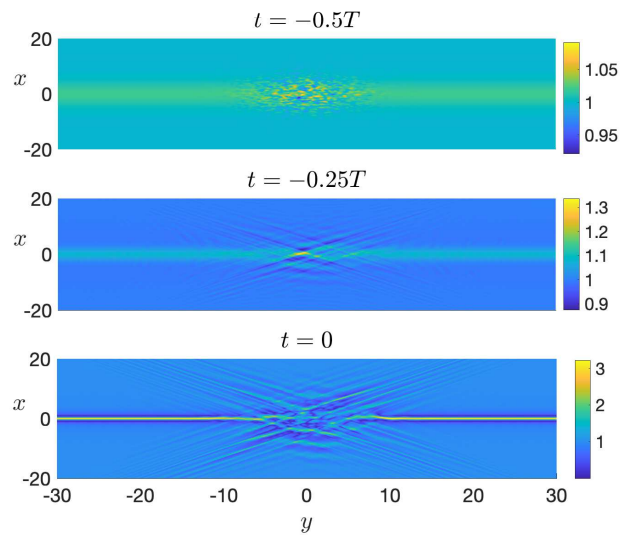}
\caption{Snapshots of {the} perturbed Peregrine soliton {($\widetilde{u} = u_{\rm P}$)} evolution {for} 
hyperbolic 
NLS ($s_1 = -1 = -s_2 $). Shown is $|u(x,y,t)|$ {seeded} with a $10\%$ localized perturbation at {$\overline{t} = - T/2$ for $T \approx 6.81$}.  \label{P_DNS_Instab}}
\end{figure}

Even though the solitons are unstable at 
large wavenumbers, it {corresponds to a suitable}
combination  of wavenumbers, namely, it must be those Fourier modes in the hyperbolic {instability} 
region shown in Fig.~\ref{MI_stab_regions}(b). 
Applying a perturbation with no $x$-dependence ($k = 0$ modes) does {\it not} yield instability (see {\cite{SuppSims}}). 
This is a subtle difference from other types of transverse instability. 

{In further support of our analytical expectations that} the stability of the Peregrine soliton is well-approximated by a limiting KM soliton, we perform {the following} simulation{, highlighted in Fig.~\ref{P_DNS_Instab}}. Well before the maximum peak occurs, the Peregrine soliton is perturbed by  the form given in (\ref{IC_perturb}). At the maximum focusing point, the line soliton has again { broken apart}
near the {region where the perturbation is applied.} We point out that perturbing the soliton earlier will result in earlier onset of the instability and degradation of the mode.

As a final note, these results for the hyperbolic 
NLS  differ immensely from those of elliptic-focusing NLS.  KM solitons  are transversely unstable  at long wavelength perturbations in the elliptic-focusing NLS equation and, 
{also} importantly, perturbed solutions can collapse in finite time. Indeed, perturbations of the sort in (\ref{IC_perturb}) {indicate blow up occurs.} 
Numerical simulations illustrating this are given in 
{\cite{SuppSims}}.  }





{In conclusion,} { transverse stability of rogue waves was studied in the 
{NLS} equation. {Linear}  stability of the Kuznetsov-Ma soliton was computed {via} Floquet theory. Since the {Kuznetsov-Ma} 
soliton approaches the  Peregrine soliton, 
{it is anticipated {that} its instability features will} also approach {those} 
of  Peregrine. {Indeed, the stability} of the Peregrine soliton was found to coincide with that of plane waves. Importantly, {in the hyperbolic case} this leads to instability at arbitrarily high frequencies from wavenumbers in {a} hyperbolic region in the spectral plane. Direct numerical solutions confirm that appropriately perturbed solitons are transversely {unstable.}
{Based on the above considerations it is natural to expect that  two dimensional {perturbations} 
can annihilate the  KM or Peregrine soliton solutions. 



\section*{Acknowledgements}
\label{Acknowledge}
This work was partially supported by AFOSR under Grant
No. FA9550-19-1-0084 and NSF under {Grants  DMS--1712793 and} DMS-2005343.

\end{document}